\documentclass[slac_one]{revtex4}
\usepackage{graphicx}
\usepackage{fancyhdr}
\begin{document}

%Title of paper
\title{{\small{2005 International Linear Collider Workshop - Stanford,
U.S.A. \hfill SLAC-PUB-11419}}\\ %% Please keep this conference title here
\vspace{12pt}
Higgs bosons in Left-Right symmetric Randall-Sundrum models} %% Paper title goes here

\author{B. Lillie}
\affiliation{SLAC, Stanford, CA 94025, USA}

\begin{abstract}

We study the corrections to Higgs physics in a model of a single
warped extra dimension with all fields except the Higgs in the bulk,
and a gauge symmetry extended to $SU(2)_L\times SU(2)_R \times
U(1)_{B-L}$. We find that generically the Higgs coupling to
electroweak gauge boson pairs is suppressed, the coupling to gluons
is enhanced, and the coupling to photons is often suppressed, but
can be enhanced.

\end{abstract}

%\maketitle must follow title, authors, abstract
\maketitle

\thispagestyle{fancy}

% body of paper here - Use proper section commands
% References should be done using the \cite, \ref, and \label commands
% Put \label in argument of \section for cross-referencing
%\section{\label{}}

\section{Randall-Sundrum in the bulk}
\footnote{Notes for a talk given at the 2005 International Linear
Collider Workshop, based on the work in \cite{Lillie:2005pt}.} One
of the most promising solutions to the hierarchy problem is the
Randall-Sundrum (RS) model \cite{Randall:1999ee}. In this model
there is a single extra dimension compactified on $S^1/Z_2$ with the
non-factorizable metric of $AdS_5$
\begin{equation}
ds^2 = \left(\frac{R}{z}\right)^2(d x^2-dz^2).
\end{equation}
Here $z$ is the extra-dimensional coordinate, and $R$ is the inverse
of the $AdS$ curvature, $k$. Two branes define the boundaries of the
extra dimension. One, at $z=R$, is called the UV, or Planck, brane.
The other, at $z=R'$, is the IR, or TeV, brane. Picking $R' =
(M_{\rm Planck}/M_{\rm electroweak})R$, which is natural in
realistic stabilization mechanisms, solves the hierarchy problem
\cite{Goldberger:1999uk}.

In the original RS model the SM was confined to the IR brane, and
only gravity propagated in the bulk \cite{Davoudiasl:1999jd}. It has
since been realized that both gauge and fermion fields can live in
the bulk in a realistic model \cite{Davoudiasl:1999tf}. These models
are realistic, but the parameter space can be strongly reduced by
precision electroweak constraints. Much of this problem can be
traced to the fact that the massive gauge fields receive a
contribution to their mass from the bulk geometry which does not
respect the custodial $SU(2)_c$. This can be fixed by expanding the
gauge group to $SU(2)_L \times SU(2)_R \times U(1)_{B-L}$, which
dramatically improves the electroweak fit \cite{Agashe:2003zs}. The
breaking of this extended electroweak symmetry proceeds in two
stages: on the UV brane $SU(2)_R \times U(1)_{B-L} \rightarrow
U(1)_Y$; on the IR brane $SU(2)_L \times SU(2)_R \rightarrow
SU(2)_D$, where $SU(2)_D$ is the diagonal of the $SU(2)$ groups.
This paper investigates the properties of the Higgs sector that
accomplishes this breaking.

We now ask what drives the breaking on each brane. On the planck
brane all degrees of freedom will have Planck scale masses, so we
can ignore them. We can then implement the breaking with boundary
conditions to good approximation. This leads to the boundary
conditions at $z=R$
\begin{eqnarray}
 \partial_z \left(\frac{\kappa}{\lambda}A_R - A_B\right) & = 0, & \partial_z A_L = 0,\hfill\nonumber \\
 A_B - \frac{\kappa}{\lambda}A_R^3 & = 0, & A_R^{\pm} =
 0.\label{eq:gaugebcR}
\end{eqnarray}
Here $\kappa$ and $\lambda$ are ratios of 5D gauge couplings:
$\kappa = g_{5R}/g_{5L}$, and $\lambda = g_{5B}/g_{5L}$.

On the TeV brane, the masses will be TeV scale, so we should look at
the Higgs sector in detail. The simplest structure that will create
the breaking pattern is a real Higgs that is a bidoublet under
$SU(2)_L \times SU(2)_R$. This leads to the boundary conditions at
$z=R'$
\begin{eqnarray}
 \partial_z (A_L + \kappa A_R) & = & 0, \hspace{1cm} \partial_z A_B = 0,\hfill\nonumber\\
 \partial_z(\kappa A_L - A_R) & = & -\frac{g^2_{5L}v^2}{4}(\kappa A_L - A_R).
 \label{eq:gaugebcRp}
\end{eqnarray}
Note that in the $v/k\to\infty$ limit we obtain the usual Higgsless
boundary conditions, and this model reduces to the Higgsless model
in \cite{Csaki:2003dt}. We will use this parameter, $v/k$ to
interpolate between the SM limit ($v/k \rightarrow 0$), and the
Higgsless limit.

\begin{figure}[t]{
 \includegraphics[angle=-90,width=12cm]{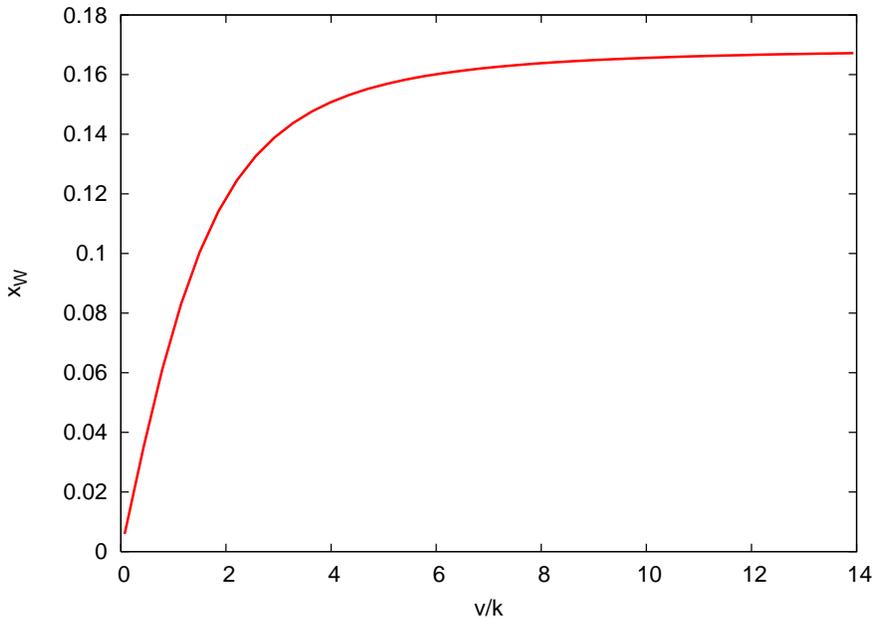}
\label{fig:wroot} \caption{Behavior of the first charged boson mass
(corresponding to the observed $W$) as a function of v/k at fixed k.
The linear behavior at small $v/k$ corresponds to the ordinary
Higgsed model limit, and the flat behavior as $v/k \to \infty$ to
the Higgsless limit.}}
\end{figure}

To write down the effective 4D theory, we expand the 5D fields into
Kaluza Klein (KK) fields,
\begin{equation}
A(x,z) = \sum_n \zeta^{(n)}_A(z)A^{(n)}(x)
\end{equation}
We can now obtain the gauge boson wavefunctions by solving the
equation of motion subject to the boundary conditions
(\ref{eq:gaugebcR}) and (\ref{eq:gaugebcRp}). This produces a
spectrum of eigenvalues corresponding to the excitations of the
gauge fields. The lowest masses in each of the charged and neutral
sectors will correspond to the $W$ and $Z$ bosons. The neutral
sector also contains a zero mode, corresponding to the photon. Fig
\ref{fig:wroot} shows the eigenvalue for the $W$ as a function of
the parameter $v/k$.

\begin{figure}[t]{
\includegraphics[angle=-90,width=12cm]{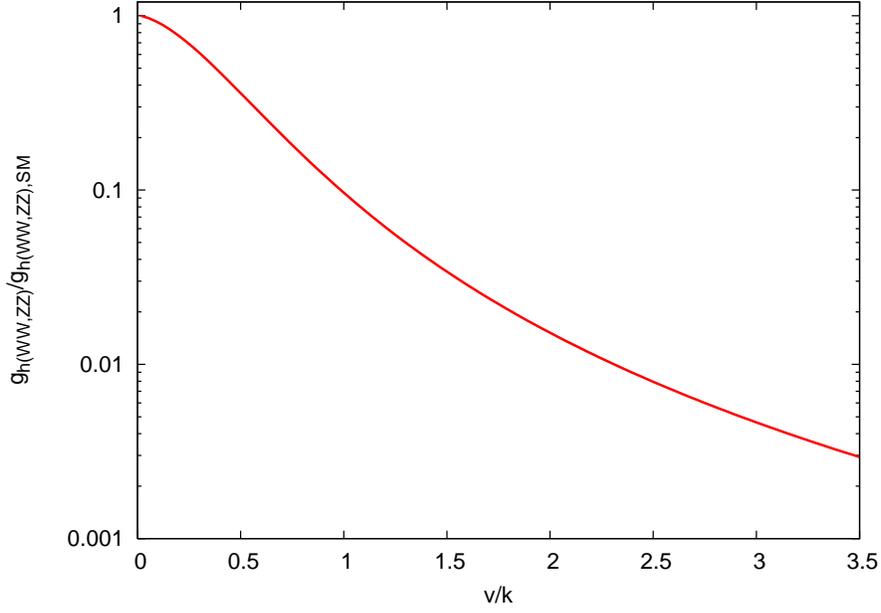}
\label{fig:wcoup} \caption{Coupling of the Higgs to vector boson
pairs compared to the SM value as a function of $v/k$. Again, the
$W$ and $Z$ coupling strengths are nearly identical due to the
custodial symmetry.}}
\end{figure}

\section{Higgs properties}

One interesting feature of this model is that the $W$ and $Z$
wavefunctions are suppressed near the IR brane, as can be seen by
inspecting the boundary conditions. This suppression increases for
increasing $v/k$. This means that the coupling of massive gauge
bosons to the Higgs will generically be suppressed. Fig.
\ref{fig:wcoup} shows the coupling of the $W$ to the Higgs. For
values of $v/k$ near unity the LEP bounds on the Higgs mass can be
dramatically reduced. (For larger values of $v/k$ the model is
effectively Higgsless.)

The fermion sector of this model is more complicated. Again, the
Higgs vev induces mixed boundary conditions that link left and right
handed fields to give the fermions masses. However, there are two
new degrees of freedom. First, since 5D fermions are achiral, ther
can always be a mass term in the bulk $m\bar \Psi \Psi$. The main
effect of this term is to shift the location of the fermion zero
mode in the bulk. By changing this parameter we can cause the zero
mode to be localized either near the UV or IR brane, and also can
change the degree of this localization. This allows us to control
the overlap of the zero mode with the IR brane, and consequently the
strength with which the fermion interacts with the Higgs. In this
way the hierarchy of fermion masses can be generated by order one
changes in the 5D masses. The second complication arises from the
$SU(2)_R$ symmetry which enforces that, for example, $m_t = m_b$ if
unbroken. This mass relation can be modified by mixing with new
fermions localized to the Planck brane, where the $SU(2)_R$ is
broken. For full details, see \cite{Csaki:2003sh}.

Note that there are tree-level corrections to precision electroweak
observables, coming largely from the KK excitations of the gauge
bosons. Unfortunately, the magnitude of these corrections is highly
sensitive to the configuration of the fermion sector. For the
specific configuration studied in \cite{Lillie:2005pt} we find the
constraint $v/k \le 1/4$. There are, however, special points in the
fermion parameter space where the constraint becomes trivial, so a
wide range of $v/k$ should be considered.

\begin{figure}[t]{
\includegraphics[angle=-90,width=12cm]{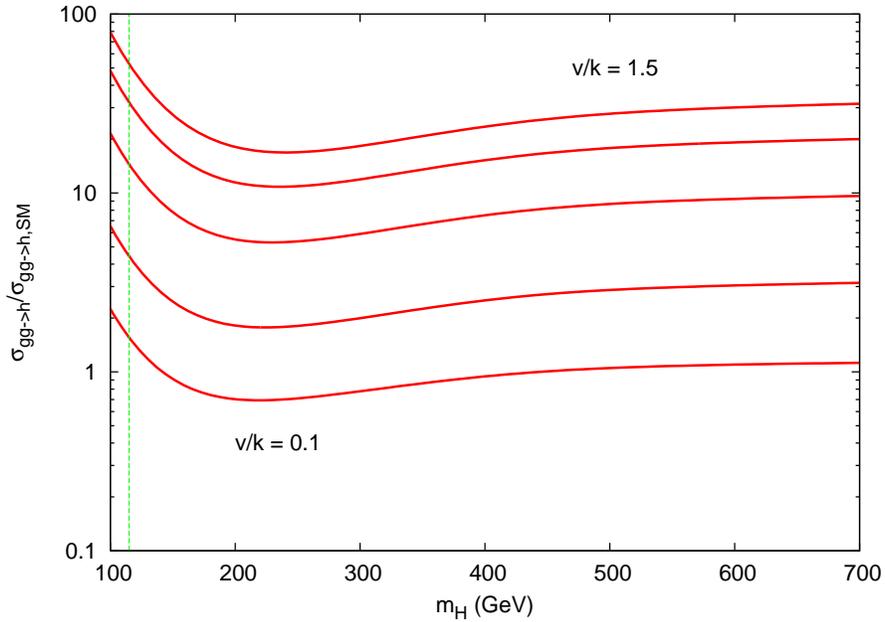}
\label{fig:hgg} \caption{Ratio (at lowest order) of the production
cross section for $gg\to h$ compared with the value in the SM. The
different curves correspond to the values of $v/k$ from top to
bottom of (1.5, 1.1, 0.8, 0.4, 0.1).}}
\end{figure}

The final interesting shift in Higgs properties is in the couplings
to massless gauge bosons, {\it i.e.} gluons and photons. The
coupling of the Higgs to gluon pairs is induced through top loops.
However, in this model the KK excitations also couple to the Higgs.
Furthermore, note that we have arranged small 4D Yukawa couplings
for the other fermions by small wavefunction overlaps with the
would-be zero modes. The 5D Yukawa couplings are all order 1, and
the excited states have {\it no} wavefunction suppression. Hence
there are large contributions to the Higgs-glue-glue coupling from
the KK excitations of {\it all} colored fermions. This leads to an
enhancement in that coupling, as seen in Fig. \ref{fig:hgg}. There
are similar corrections to the Higgs-gamma-gamma coupling. The
situation there is more complicated, however, since there are also
contributions from $W$ boson loops, which are dominant in the SM,
and the Higgs coupling to $W$s is suppressed.

\begin{figure}[t]{
\includegraphics[angle=-90,width=12cm]{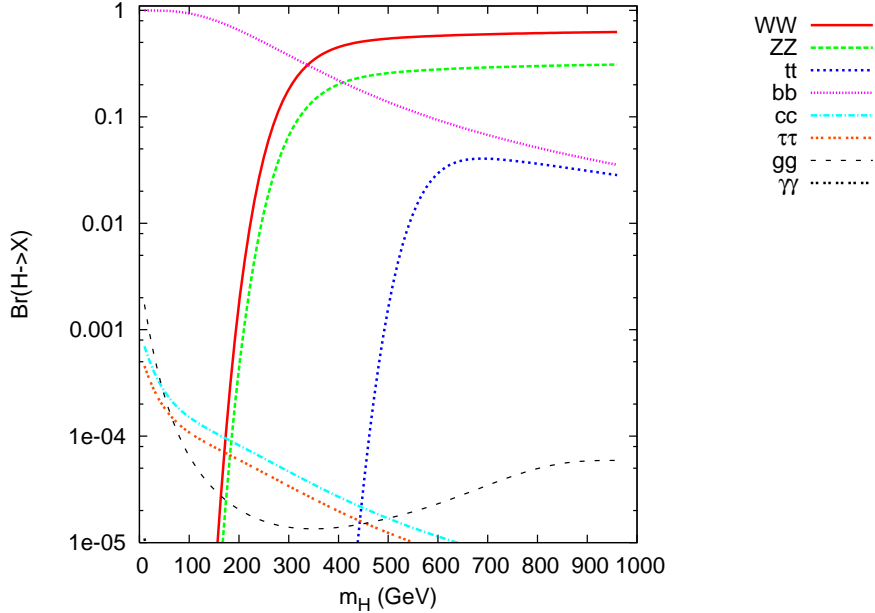}
\label{fig:widthbr} \caption{Branching ratios for Higgs decay into
various channels as a function of the Higgs mass at fixed $v/k =
1/10$.}}
\end{figure}

We can now look at the behavior of the Higgs branching ratios, as
shown in Fig. \ref{fig:widthbr}. Note in particular the dominance of
$h\rightarrow b \bar b$ over a wide range, and the late onset of
$h\rightarrow (WW,ZZ)$. This is driven by the $SU(2)_R$ symmetry,
which gives a large enhancement of the $b$-quark Yukawa, and the
suppression of the gauge boson couplings. Note also the reduction in
the $h\rightarrow \gamma\gamma$ mode. This will make discovery at
the LHC difficult. The suppression of the coupling to gauge bosons
also means that production at the ILC will be reduced, making this
Higgs a particularly difficult one to find. However, it will be
essential to measure the Higgs couplings with precision to identify
a warped extra dimension as the correct theory of new physics, if
indeed this is what is realized in nature.

% If you have acknowledgments, this puts in the proper section head.
\begin{acknowledgments}
Work supported by Department of Energy contract DE-AC02-76SF00515.
\end{acknowledgments}

\end{document}